\documentclass{aa}

\usepackage[varg]{txfonts}
\usepackage{graphicx}
\usepackage{natbib}

\begin{document}

\title{Time-series photometry of Earth flyby asteroid 2012~DA$_{14}$}

\author{Tsuyoshi Terai\inst{1}
  \and Seitaro Urakawa\inst{2}
  \and Jun Takahashi\inst{3}
  \and Fumi Yoshida\inst{1}
  \and Goichi Oshima\inst{4}
  \and Kenta Aratani\inst{4}
  \and Hisaki Hoshi\inst{4}
  \and Taiki Sato\inst{4}
  \and Kazutoshi Ushioda\inst{4}
  \and Yumiko Oasa\inst{4}
}

\institute{National Astronomical Observatory of Japan, 2-21-1 Osawa, Mitaka, Tokyo 181-8588, Japan\\
  \email{tsuyoshi.terai@nao.ac.jp}
  \and Bisei Spaceguard Center, Japan Spaceguard Association, 1716-3 Okura, Bisei, Ibara, Okayama
       714-1411, Japan
  \and Center for Astronomy, University of Hyogo, 407-2 Nishigaichi, Sayo-cho, Sayo-gun, Hyogo
       679-5313, Japan
  \and Faculty of Education, Saitama University, 255 Shimo-Okubo, Sakura, Saitama 338-8570, Japan
}

\date{Received 28 June 2013 / Accepted 12 September 2013}

\abstract 
{The object 2012 DA$_{14}$ is a near-Earth asteroid with a size of several tens of meters.
It had approached closely the Earth on 15 February, 2013 UT, providing an opportunity for precise
measurements of this tiny asteroid.}
{The solar phase angle of 2012 DA$_{14}$ had varied widely around its closest approach but was
almost constant during the following night.
We performed time-series photometric observations on those two nights to determine the rotational
properties and phase effect.}
{The observations were carried out using the 0.55-m telescope at Saitama University, Japan.
The $R$-band images were obtained continuously over a 2 hr period at the closest approach and for
about 5 hr on the next night.}
{The lightcurve data from the second night indicates a rotational period of
11.0$^{+1.8}_{-0.6}$ hr and a peak-to-peak amplitude of 1.59 $\pm$ 0.02 mag.
The brightness variation before and after the closest approach was separated into two components
that are derived from the rotation and phase effect.
We found that the phase curve slope of this asteroid is significantly shallower than those of
other L-type asteroids.}
{We suggest that 2012 DA$_{14}$ is coated with a coarse surface that lacks fine regolith particles
and/or a high albedo surface.}


\titlerunning{Time-series photometry of 2012 DA$_{14}$}
\authorrunning{Terai et al.}

\maketitle

%%%%%%%%%%%%%%%%%%%%%%%%%%%%%%%%%%%%%%%%%%%%%%%%%%%%%%%%%%%%%%%%%%%%%%%%%%%%%%%%%%%%%%%%%%%%%%%%%%%
\section{Introduction}     \label{sec:introduction}  %%%%%%%%%%%%%%%%%%%%%%%%%%%%%%%%%%%%%%%%%%%%%%
%%%%%%%%%%%%%%%%%%%%%%%%%%%%%%%%%%%%%%%%%%%%%%%%%%%%%%%%%%%%%%%%%%%%%%%%%%%%%%%%%%%%%%%%%%%%%%%%%%%

On February 15, 2013 UT, the near-Earth object (NEO), 2012~DA$_{14}$, passed closely to the Earth
at a distance of about 27,700~km inside a geosynchronous orbit \citep{2012MNRAS.427.1175W}.
Its diameter was estimated to be probably less than 50~m from the Goldstone radar
measurements$\footnote{Presented by Lance A. M. Benner at
\url{http://echo.jpl.nasa.gov/asteroids/2012DA14/2012DA14_planning.html}}$.
Since ten-meter sized asteroids are too faint to be observed in detail, their population, internal
structure, and surface properties remain uncertain even though these are required for estimating
the frequency and influence of hazardous impact events onto the Earth
\citep[e.g.,][]{2002aste.conf..739M}.
These small asteroids also have short dynamical/collisional lifetimes and a feeble surface gravity,
so that their surface properties and structure could be substantially different from
1--100~km-sized asteroids.

The Earth flyby made 2012~DA$_{14}$ bright enough to be precisely observed by relatively small
telescopes on the ground.
This event was an exciting opportunity to investigate the surface properties of such a small
asteroid.
Its visible/near-infrared colors and visible spectra were already presented by
\citet{2013A&A...555L...2D} and \citet{2013arXiv1306.2111U}.
Interestingly, both studies classified 2012~DA$_{14}$ as an L-type, a minor taxonomic class
among asteroid population.
De~Le\'on et al. (2013) also indicated the rotational period of 8.95~$\pm$~0.08~hr.
In this paper, we focus on the relationship between reflectance and solar phase angle, which is
known as the photometric phase curve.
It provides useful indications of the surface properties, such as geometric albedo and regolith
structure.
We present the rotational lightcurve and phase curve of 2012~DA$_{14}$ from our time-series photometric
observations around the closest approach and on the next night.

%%%%%%%%%%%%%%%%%%%%%%%%%%%%%%%%%%%%%%%%%%%%%%%%%%%%%%%%%%%%%%%%%%%%%%%%%%%%%%%%%%%%%%%%%%%%%%%%%%%
\section{Observations and measurements}     \label{sec:observations}        %%%%%%%%%%%%%%%%%%%%%%%
%%%%%%%%%%%%%%%%%%%%%%%%%%%%%%%%%%%%%%%%%%%%%%%%%%%%%%%%%%%%%%%%%%%%%%%%%%%%%%%%%%%%%%%%%%%%%%%%%%%

Our observations have been carried out on the two consecutive nights of February 15 and 16, 2013
(UT).
We used the 0.55-m telescope at Saitama University in Saitama, Japan
(139.6059$\degr$~E, 35.8624$\degr$~N) with the FLI Micro Line ML-4710 CCD
(1056~$\times$~1027 pixels) mounted on the prime focus that covers a
32$\arcmin$~$\times$~32$\arcmin$ field of view (FOV).
All images have been taken with sidereal tracking.
Before the target crossed the edge of the FOV, we moved the telescope in the same direction that
the asteroid moved so that the asteroid would stay in the FOV.
Most of the imaging data were continuously obtained over the observations with the Johnson
$R$-band.
The typical seeing was 3$\arcsec$--4$\arcsec$.

\begin{figure}
\resizebox{\hsize}{!}{\includegraphics{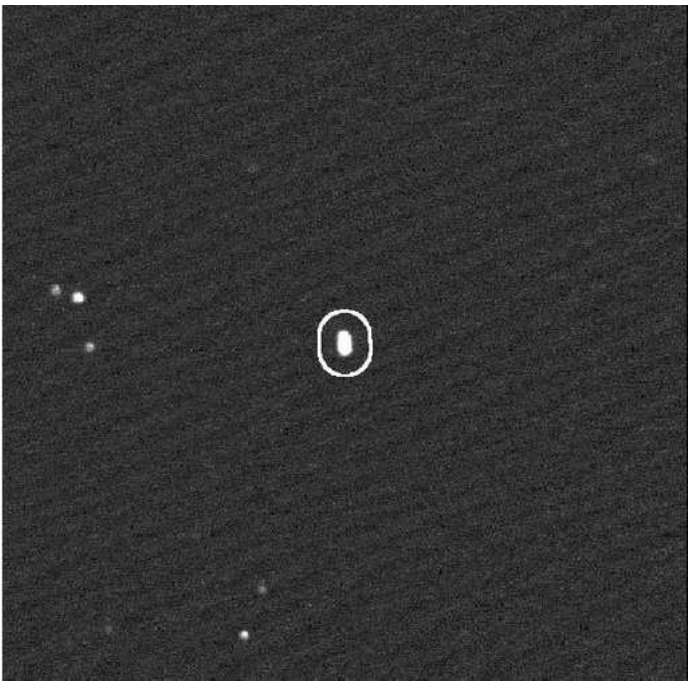}}
\caption{
An $R$-band image of 2012~DA$_{14}$ taken by the 0.55-m Saitama University telescope at the closest
approach with a 0.5-sec exposure.
The field-of-view covers 20$\arcmin$~$\times$~20$\arcmin$.
The solid line represents the moving-circular aperture used for photometry.
}
\label{fig:1}
\end{figure}

In the first night, the time-series imaging was performed around the closest approach of
2012~DA$_{14}$ at 19:33~UT over a period of $\sim$2~hr.
The asteroid 2012~DA$_{14}$ moved from the altitude of 18$\degr$ to 40$\degr$ with a sky motion of
20--52~arcmin~min$^{-1}$.
The solar phase angle decreased from 38$\degr$ to 19$\degr$ before 19:55~UT, then increased until
42$\degr$.
We obtained more than 2,000 images that were taken every 2~sec with a 0.5-sec exposure by using
defocus imaging to avoid saturation of the target asteroid.
All the images were bias corrected, flat-fielded, and sky subtracted.
In spite of the short exposure, the shape of 2012~DA$_{14}$ images is elongated due to its extremely
fast sky motion.
Its barycenters may not coincide with the position that coincides with the center of the exposures.
This discrepancy could cause an uncertainty of the measured central coordinates.
For improving the positioning accuracy, pixel values for light sources on each image were replaced
by a constant value substantially that is larger than the sky level.
The central coordinates of the asteroid were then determined by ellipse fitting of the processed
images.
The total flux of the asteroid on each image was calculated based on the total counts within an
elongated circular aperture with a radius of 30$\arcsec$ assuming the synthetic shape of the
asteroid image (see Figure~\ref{fig:1}).
If a field star overlaps with 2012~DA$_{14}$, the data were excluded.
Using the field stars identified as USNO-B1.0 stars with an $R$~$<$~12~mag
\citep{2003AJ....125..984M}, the zero magnitude point and atmospheric extinction coefficient were
estimated.

%%%%% Figure 1 %%%%%

During the second night, the identical observation was conducted over $\sim$5~hr.
However, the exposure times were 10--60~sec because the asteroid became faint ($R$~$\sim$~15~mag)
and slow ($<$~0.3~arcmin~min$^{-1}$) compared to the previous night.
The images were corrected with the same reduction processes as the first night data.
The brightness variation of 2012~DA$_{14}$ was measured with relative photometry using ten or more
field stars in every sky area.
The aperture was the same as that mentioned above but with a radius of 10$\arcsec$.
The Landolt standard stars observed immediately before this photometric sequence were used for
flux calibration.

%%%%%%%%%%%%%%%%%%%%%%%%%%%%%%%%%%%%%%%%%%%%%%%%%%%%%%%%%%%%%%%%%%%%%%%%%%%%%%%%%%%%%%%%%%%%%%%%%%%
\section{Results}     \label{sec:results}            %%%%%%%%%%%%%%%%%%%%%%%%%%%%%%%%%%%%%%%%%%%%%%
%%%%%%%%%%%%%%%%%%%%%%%%%%%%%%%%%%%%%%%%%%%%%%%%%%%%%%%%%%%%%%%%%%%%%%%%%%%%%%%%%%%%%%%%%%%%%%%%%%%

\begin{figure}
\resizebox{\hsize}{!}{\includegraphics{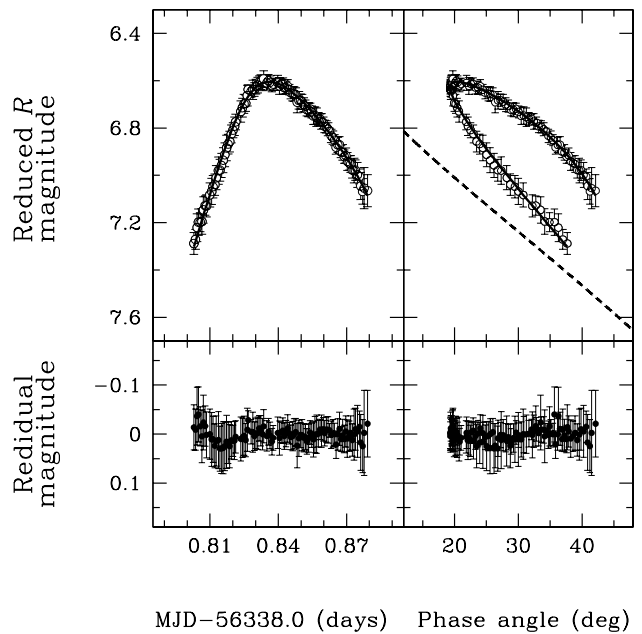}}
\caption{
The brightness variations of 2012~DA$_{14}$ around the closest approach at 19:33 UT on February 15,
2013 (MJD 56338.8146).
The upper left and upper right panels show the lightcurve with time and solar phase angle,
respectively.
The open circles and solid line represent the photometric data and the best-fit model generated by
a combination of the H-G phase function.
The dashed line in the upper right panel represents the model phase function with $G$~=~0.44, which
was derived from our best-fit model.
The lower left and lower right panels show the residual to the best-fit model of the lightcurves
with time and solar phase angle, respectively.
In the all panels, the error bars indicate the 1~$\sigma$ photometric uncertainties.
}
\label{fig:2}
\end{figure}

The results of photometry in the first night are shown in Figure~\ref{fig:2} as lightcurves with
respect to time and phase angle.
The data points and error bars represent the averaged magnitude over every minute and its standard
deviation, respectively.
The mean uncertainty is 0.04~mag.
The magnitude was adjusted for the heliocentric and geocentric distances to those at the closest
approach to the Earth.
The brightness reached a peak at the minimal phase angle, indicating that the lightcurve was
dominated by the phase effect.
However, there is a discrepancy between the phase curves before and after the peak.
This is likely to be due to the rotational variation in brightness, as discussed in
Section~\ref{sec:discussion}.

%%%%% Figure 2 %%%%%

\begin{figure}
\resizebox{\hsize}{!}{\includegraphics{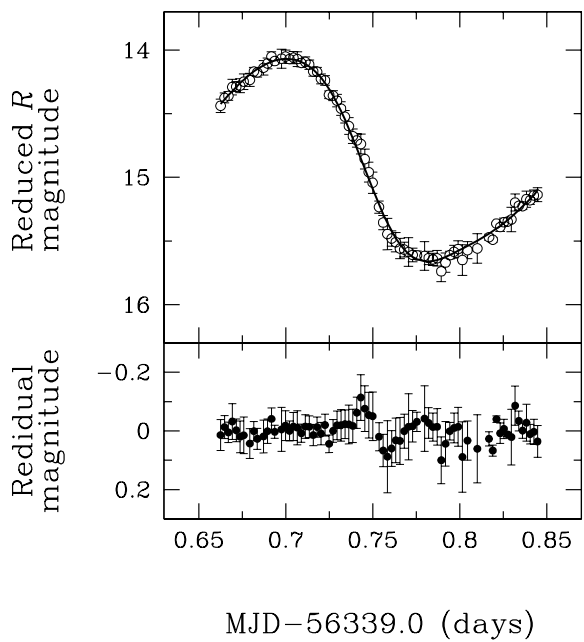}}
\caption{
The obtained lightcurve of 2012~DA$_{14}$ on February 16, 2013 UT.
The upper panel shows the photometric data (open circles) and the best-fit
model with a Fourier series curve (solid line).
The lower panel shows the residual of the best-fit model.
In the both panels, the error bars indicate the 1$\sigma$ photometric uncertainties.
}
\label{fig:3}
\end{figure}

\begin{figure}
\resizebox{\hsize}{!}{\includegraphics{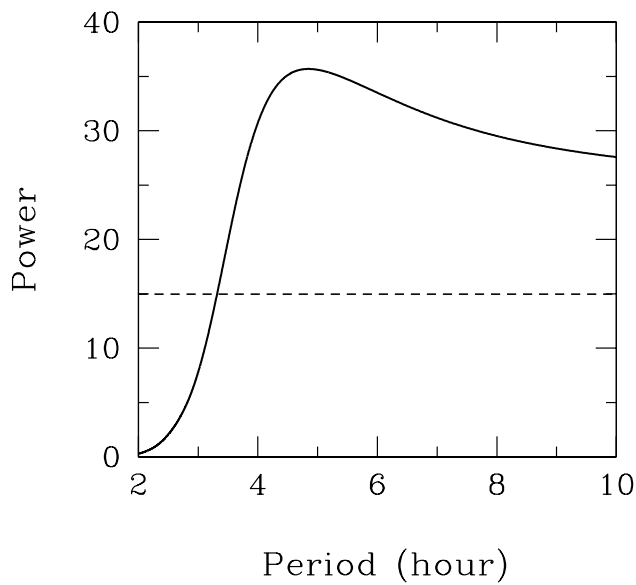}}
\caption{
Periodogram of the photometric data for 2012~DA$_{14}$ obtained at February 16, 2013 UT.
The dashed line shows the 99.99\% confidence level.
}
\label{fig:4}
\end{figure}

Figure~\ref{fig:3} shows the lightcurve from the second night data.
The data points and error bars represent the averaged magnitude over every 3~min and its standard
deviation, respectively.
The mean uncertainty is 0.05~mag.
The magnitude was adjusted for the heliocentric and geocentric distances to those at the beginning
of the observation.
Because the variation in phase angle is negligible over the observation time of the second night
(82.2$\degr$--82.6$\degr$), the lightcurve shows the brightness change caused by the asteroid's
rotation.
We have analysed the lightcurve using the Lomb-Scargle periodogram technique
\citep{1976Ap&SS..39..447L,1982ApJ...263..835S}.
Figure~\ref{fig:4} shows the resulting periodogram that indicates the plausible apparent period of
around 5~hr.
The synthetic curve was determined by fitting the forth-order Fourier series formulation.
The best-fit model gives an apparent period of 5.5$^{+0.9}_{-0.3}$~hr with a peak-to-peak
amplitude of 1.59~$\pm$~0.02~mag, where the uncertainties correspond to 1$\sigma$.
Assuming a double-peaked lightcurve, we obtained the rotational period of
11.0$^{+1.8}_{-0.6}$~hr.
The best-fit model and its residual are shown in Figure~\ref{fig:3}.

%%%%% Figure 3 %%%%%

%%%%% Figure 4 %%%%%

De~Le\'on et al. (2013) presented the rotational lightcurve of 2012~DA$_{14}$ covering from
56338.85 to 56339.20 in modified Julian date (MJD), which has some discontinuities and sharp turns.
They obtained a rotational period of 8.95~$\pm$~0.08~hr.
The shape of our lightcurves and around the second maximum at MJD~56339.12 in
\citet{2013A&A...555L...2D} match each other.
However, we could not find the rotation period of 8.95~hr in the combined lightcurve that was
derived from our observations and those of de~Le\'on et al..
We note that our lightcurve was acquired when the asteroid was moving slowly and the solar phase
angle was almost constant, or in good condition for precise measurement of the rotational
brightness variation without the phase effect.
During de~Le\'on et al.'s observation on February 15, however, 2012~DA$_{14}$ still had a rapid sky
motion (1--20~arcmin~min$^{-1}$) and a significant change in the phase angle
(50$\degr$--75$\degr$).
These observational difficulties might have caused the inconsistent lightcurve with our result.
Another possibility is that the asteroid might have experienced a significant spin down in the time
during the close encounter with Earth.
We adopted 11.0$^{+1.8}_{-0.6}$~hr as the most probable rotational period of 2012~DA$_{14}$,
although shorter periods cannot be ruled out.

%%%%%%%%%%%%%%%%%%%%%%%%%%%%%%%%%%%%%%%%%%%%%%%%%%%%%%%%%%%%%%%%%%%%%%%%%%%%%%%%%%%%%%%%%%%%%%%%%%%
\section{Discussion}     \label{sec:discussion}      %%%%%%%%%%%%%%%%%%%%%%%%%%%%%%%%%%%%%%%%%%%%%%
%%%%%%%%%%%%%%%%%%%%%%%%%%%%%%%%%%%%%%%%%%%%%%%%%%%%%%%%%%%%%%%%%%%%%%%%%%%%%%%%%%%%%%%%%%%%%%%%%%%

\subsection{Model fitting}

To understand the brightness variation of 2012~DA$_{14}$ around its closest approach, as seen in
Figure~\ref{fig:2}, we use synthetic models generated from a combination of a given phase
function and a rotational lightcurve.
The latter was based on the measurements from the data of the second night assuming no variation of
the rotational period from the time around the closest approach to the second night.
We note that the interval between the centered times of observations at the first and second nights
are about 22~hr in length corresponding to almost exactly two rotations; that is, the surface on
the similar side was observed on the both nights.
The rotational period and lightcurve pattern were determined from the second night data, but the
peak-to-peak amplitude was parameterized.
This is because the difference in inclination of the spin axis to the line of sight between the two
nights is unknown, and the brightness amplitude of an ellipsoidal body depends on the phase angle
\citep{1989aste.conf..557H}.

Model fitting was performed with the following three types of empirical phase functions using
chi-square minimization:
(i) a linearly decreasing function expressed as $b \alpha$ where $b$ is the linear coefficient
in mag~deg$^{-1}$ and $\alpha$ is the solar phase angle
\citep{1996A&AS..115..475S, 2003Icar..161...34K},
(ii) the so-called H,~G phase function with the slope parameter $G$ \citep{1989aste.conf..524B}, and 
(iii) the H,~G$_{12}$ function developed by \citet{2010Icar..209..542M}.
In the first function, the non-linear (exponential) brightness increase at small phase angles
($\alpha$~$<$~10$\degr$), known as the opposition effect, was ignored because of a lack of data.
The slope parameter ($b$, $G$, or $G_{12}$), absolute magnitude, amplitude of the rotational variation,
and $\chi^2$ value of the best-fit model with each of the phase functions are listed in
Table~\ref{table:1}.
We obtained nearly identical phase curves with those three functions though the H,~G phase function
matched with the observation most closely.
The best-fit model is shown in Figure~\ref{fig:2}.
The obtained slope parameter is $G$~=~0.44$^{+0.06}_{-0.08}$.

%%%%% Table 1 %%%%%
\begin{table}
\caption{Parameters of the best-fit model for the lightcurve of 2012~DA$_{14}$ around its closest
approach.}
\label{table:1}
\centering
\renewcommand{\arraystretch}{1.5}
\begin{tabular}{c c c c c}
\hline\hline
Phase    & Slope     & $H_R$\tablefootmark{a} & Amplitude\tablefootmark{b} & $\chi^2$ \\
function & parameter & (mag)                  & (mag)                      &          \\
\hline
Linear      & $b$=0.022$^{+0.002}_{-0.001}$    & 25.02$^{+0.07}_{-0.08}$ \tablefootmark{c} & 1.00$\pm$0.04 & 12.6 \\
H, G \ \ \  & $G$=0.44$^{+0.06}_{-0.08}$ \ \ \ & 24.78$^{+0.11}_{-0.09}$ \ \               & 1.02$\pm$0.02 & 12.2 \\
H, G$_{12}$ & $G_{12}$=0.05$^{+0.02}_{-0.02}$  & 24.64$^{+0.08}_{-0.08}$ \ \               & 0.90$\pm$0.05 & 18.3 \\
\hline
\end{tabular}
\tablefoot{
\tablefoottext{a}{Absolute magnitude in the $R$-band.}
\tablefoottext{b}{Peak-to-peak amplitude of the rotational lightcurve.}
\tablefoottext{c}{This value has been overestimated because the non-linear brightness increase at
small phase angles was ignored.}
}
\end{table}

It is widely known that asteroid phase curves are tied to surface properties and differ for
different taxonomic classes \citep{2002aste.conf..123M}.
Specifically, the slope of the phase curve in the linear part
(10$\degr$~$\lesssim$~$\alpha$~$\lesssim$~50$\degr$) is inversely correlated with the geometric
albedo \citep{1996A&AS..115..475S,2000Icar..147...94B}.
The asteroid 2012~DA$_{14}$ has been classified as an L-type asteroid
\citep{2013A&A...555L...2D,2013arXiv1306.2111U}.
The mean value of the geometric albedo ($p_V$) for L-type asteroids is $p_V$~=~0.14~$\pm$~0.04 by
AKARI \citep{2013ApJ...762...56U} or 0.18~$\pm$~0.08 by WISE \citep{2011ApJ...741...90M}, which is
lower than that of S-type asteroids ($p_V$~=~0.23~$\pm$~0.07).
This indicates that L-type asteroids have steeper phase curve slopes (i.e. lower $G$ value) than
S-type asteroids regardless of the surface composition.
The mean $G$ value of S-type asteroids is 0.23~$\pm$~0.02 \citep{1990A&AS...86..119L}.
Actually, the L-type asteroids (236) Honoria ($p_V$~=~0.11~$\pm$~0.02), (387) Aquitania
($p_V$ is unknown), and (980) Anacostia ($p_V$~=~0.14~$\pm$~0.03) show the $G$ values of
-0.020~$\pm$~0.014, 0.02~$\pm$~0.02, and 0.060~$\pm$~0.003, respectively
\citep{2012Icar..221..365P}.
The typical phase curve of L-type asteroids is still unclear but is expected to have a steeper
slope than that of S-type asteroids.
However, the phase curve of 2012~DA$_{14}$ that is obtained from the best-fit model is much
shallower than those of the known L-type asteroids and even S-type asteroids.
This fact implies that 2012~DA$_{14}$ could have peculiar surface properties compared to other
L-type asteroids.

\subsection{Interpretation}

We suggest two potential mechanisms causing the shallow phase curve of 2012~DA$_{14}$.
One is that the surface of 2012~DA$_{14}$ could be coated with coarse regolith.
These small asteroids have a weak gravitational field and have difficulty in retaining fine
particles produced by meteorite impacts.
At relatively high phase angles ($\alpha$~$\gg$~1$\degr$), asteroid phase curves are dominated by
the shadow-hiding effect rather than the coherent backscattering
\citep{2000Icar..147...94B,2007PASP..119..623F}.
Given a power-law size distribution of the regolith particles, an increase in the smallest particle
size widens an angular width parameter of the phase curve with the shadow-hiding effect
\citep{1993tres.book.....H}.
A lack of fine particulate regolith provides a reasonable explanation for the shallower phase
curve compared to larger asteroids.

Another reason is that 2012~DA$_{14}$ could have a high albedo surface.
As mentioned in the previous section, the slope of the asteroid phase curve in the intermediate
range of the solar phase angle is known to have a close correlation with geometric albedo
\citep{1996A&AS..115..475S,2000Icar..147...94B}.
E-type asteroids with a mean albedo of 0.41 \citep{2013ApJ...762...56U} have a mean $b$ value of
0.020~$\pm$~0.002~mag~deg$^{-1}$ \citep{2000Icar..147...94B} and a mean $G$ value of
0.45~$\pm$~0.03 \citep{1990A&AS...86..119L}.
The best-fit $b$ and $G$ values are consistent with those values, indicating that the albedo of
2012~DA$_{14}$ is as high as E-type asteroids.
Using the empirical correlation between the $b$ value and visible geometric albedo, as given by
\citet{2000Icar..147...94B}, the albedo of 2012~DA$_{14}$ is expected to be
$p_V$~=~0.42~$\pm$~0.03.
De~Le\'on et al. (2013) also presented a similar albedo of 2012~DA$_{14}$ with a value of
$p_V$~=~0.44~$\pm$~0.20 though the uncertainty is large.
These high albedo values of 2012~DA$_{14}$ disagree with the fact that it is a member of L-type
asteroids with a typical albedo less than 0.20.

Generally, small asteroids seem to be covered by a young surface.
These S-type asteroids have a surface with less reddening/darkening effects due to space weathering
processes.
Considering that the collisional lifetime for main-belt asteroids smaller than 50~m in diameter is
less than 10~Myr \citep{2005Icar..178..179O}, the migration from the main-belt resonances to
Earth-crossing orbits is 0.1--1~Myr, and the typical dynamical lifetime of Earth-crossing NEOs
is $\sim$1~Myr \citep{1997Sci...277..197G}, the age of 2012~DA$_{14}$ is likely to be several Myr.
In S-type asteroids, the timescale of the surface maturation is suggested to be 10$^5$--10$^6$~yr
\citep{2009Natur.458..993V,2013Icar..225..781S}.
However, it is unclear how the reddening/darkening effects proceed on the surface of L-type
asteroids.
We cannot assess the relation between the albedo and surface age for 2012~DA$_{14}$.
One possible cause we can point out is the resurfacing effect, which occurs when planetary
encounters freshen asteroid surfaces by tidal stress \citep{2010Natur.463..331B}.
The asteroid 2012~DA$_{14}$ has frequent close approaches to the Earth \citep{2012MNRAS.427.1175W}.
If this mechanism effectively acts on 2012~DA$_{14}$, its surface could keep high albedo.

In summary, we found that 2012~DA$_{14}$ rotates with at least an 11-hr period since the last Earth
flyby.
The asteroid shows a significantly shallow phase curve which is inconsistent with known L-type
asteroids.
Such an unusual phase curve indicates that 2012~DA$_{14}$ is covered with a coarse and/or bright
surface.
This result provides useful clues for understanding the origins and evolutions of asteroid surface
properties and developing Spaceguard strategies.

\begin{acknowledgements}
This study is based on data collected at Saitama University observatory.
We thank Yusuke Takahara, Daichi Takai, Aiko Enomoto, and Kaori Nakazato
for their help during our observations.
This work was supported by JSPS Grants-in-Aid for Scientific Research 25870124 
from MEXT.
\end{acknowledgements}

%%%%%%%%%%%%%%%%%%%%%%%%%%%%%%%%%%%%%%%%%%%%%%%%%%%%%%%%%%%%%%%%%%%%%%%%%%%%%%%%%%%%%%%%%%%%%%%%%%%
%% References %%%%%%%%%%%%%%%%%%%%%%%%%%%%%%%%%%%%%%%%%%%%%%%%%%%%%%%%%%%%%%%%%%%%%%%%%%%%%%%%%%%%%
%%%%%%%%%%%%%%%%%%%%%%%%%%%%%%%%%%%%%%%%%%%%%%%%%%%%%%%%%%%%%%%%%%%%%%%%%%%%%%%%%%%%%%%%%%%%%%%%%%%

\end{document}